# Temporal dynamics in acoustic emission of the organic self-propagating high-temperature synthesis.

A.L. Parakhonsky*[1] and E.G. Klimchuk[2]

[1]*Institute of Solid State Physics, Russian Academy of Sciences,142432, Chernogolovka, Moscow Region, Russia*,
[2]*142432, Institute of Structural Macrokinetics and Materials Science, Russian Academy of Sciences,142432, Chernogolovka, Moscow Region, Russia*

Nonlinear dynamics procedures as well as spectral-correlation analysis are used for the acoustical response of consecutive physical-chemical stages of the organic self-propagating high-temperature synthesis. The transient nature of processes in the system studied is revealed. The study of changes in temporal and spatial dynamics of such systems allows driving the wave of the chemical reaction to make the solid phase "combustion" more efficient and produce condensed products with desired properties.

**Acoustical emission reflects and makes clearer the processes of the beginning, developing and dissipation of the organic self-propagating high temperature synthesis (OSHS). This results in the phase portraits of corresponding physical-chemical steps of OSHS due to the transformations of the sample structure as well as formation of resulting condensed products. The transition from one stage to another is characterized by successive transformation of a phase portrait of dynamic system and multiple changes of the phase volume. The spectral-correlation analysis in combination with the analysis of correlation dimension of attractors and Kolmogorov-Sinai entropy allowed revealing transient phenomena in the system: from regular to deterministic one and in reverse order. The approach enables to find the maximum efficiency of the mixture "combustion" process for producing high quality and ecologically cleaner synthesized materials (both final and intermediate) and it can provide a basis for developing the theory having predictive force.**

______________

*Corresponding author. Tel.: (+7) 4965228231.
E-mail address: alpar@issp.ac.ru (A.L. Parakhonsky).

**Introduction**

The use of the methods developed in the theory of dynamic systems (TDS) is very fruitful when considering systems of very different natures. It concerns both subsystem interactions in the specific system and prediction of its dynamics [1]. On the basis of the behavior of a dynamic variable's trajectory in the phase space depending on initial conditions, we can classify of a system or identify of its individual portrait. It may be appropriate not only from a purely mathematical point of view since, in some ways, it is more easily to deal with mappings than with a system of differential equations. It is so also taking into account that reconstruction of attractors in some cases is the only way to define the character of a real system [2]. Important aspect of experimental studies of transition to chaos in actual physical and chemical systems is the choice of a time variable. These may be velocity areas, a temperature or temperature gradients and so on [3]. In such a case, the idea of phase space not only looks impressive (due to representation of a complex system by a single point and the time evolution by its motion in the phase space). But it is also provides an effective way to extract the desired signal from the noise and study the transitional regimes of complex systems. In some cases, dimensional characteristics of attractors allow to follow up the scenario of transition from one state of a system to another and calculating ones parameters. This approach could be useful to investigate the chemical kinetics, self-oscillations in specific chemical systems and stable cycle's existence in them [4]. The organic self-propagating high-temperature synthesis could be considered as a dynamical system evolving under some rule [5]. Taking into consideration the characteristics of sound waves associated with OSHS processes, it might be used the methods developed in TDS theory to study this system. This is the case, for example, the question whether the spatial and temporal sound wave's field pattern is regular or more complex, are there transient processes or the mixed states? In this paper, we turned to the consistent analysis of trajectories in the phase space of OSHS sound wave combined with statistical methods. Analyzing the temporal dynamics of acoustic signal of OSHS we show the role of such a metric invariant as Kolmogorov-Sinai (KS) entropy by using a set of samples.

**1. OSHS wave's physical-chemical processes and the method of acoustical emission.**

The OSHS is attributed to the class of autowave processes which are known collectively as "solid-phase combustion". It is an analogue of the method known as self-propagating high temperature synthesis (SHS) [6]. The essence of this method consists in formation of a stationary wave of exothermic chemical reaction in a stoichiometric mixture of reactive chemicals (powders, microcrystals) by means of a local chain chemical reaction excitation due to a heat pulse [7]. The OSHS processes differ favorably from conventional methods of organic synthesis, mainly because these technological schemes do not contain organic solvents. It improves substantially environmental, safety and economic indicators. The synthesized condensed materials have a unique micro- and macrostructure due to heat and mass transfer phenomena for the components during the synthesis and the resulting product formation. Second, requirements to the equipment and OSHS methodological tools are lowered significantly compared to SHS ones due to relatively low temperatures in OSHS environments (70-250°C instead of 1000°C and more). Meanwhile, some effects in OSHS systems are revealed that do not fit into the SHS theory and do not base on their own measuring techniques. Traditionally, theoretical basis in describing the inorganic SHS compounds is a thermal conductivity equation, an equation of mass transfer and an equation of chemical kinetics [6, 8]. However, this approach is not used systematically in the case of organic mixtures for several reasons. This is because that it is not always possible to determine boundary heat-exchange conditions as well as the energy transfer at heat exchange can be due to percolation mechanism. This is, for example, the case if the reaction occurs in the porous material with a fractal structure [9]. These difficulties may be related also to the macroscopic medium motion. Existing experimental techniques are found to be too laborious, or they yield little information. Above mentioned difficulties are mainly determined by the specificities of the OSHS process itself and more complicated physical and chemical processes in such a system in comparison to chemical kinetics of inorganic substances and traditional material physics. As we have mentioned above, the percolating phase transition is beginning to play an increasingly important role in heat and mass transfer [9]. This requires new special methods of measurement and experimental data processing. In the light of these circumstances, the promising method of OSHS soundwave analysis has been developed and implemented [10]. This consists in the recording and processing of acoustical signals from the empty camera (background), reactive samples and acoustic tags inserted in the mixture. Experimental setup for the acoustic measurements includes a pulsed thermal excitation source, a reaction cell for organic and low-temperature inorganic mixtures, a sound isolating camera, a low noise microphone with parabolic sound concentrator and the system for recording and processing of acoustical signals. A detailed description of the acoustic measurement technique as well as physical and chemical processes occurring in one of the OSHS systems is set out in [11]. Identification of the data during the record of acoustic signals was carried out on the basis of the known general features of the OSHS mechanism. It includes the following steps: 1) the fast reacting of components in the course of mixing; 2) the slow reaction before the initiation of the "combustion" wave; 3) the fast reaction in the wave front accompanied by low outgassing and 4) the cooling and formation of the finite micro- and macrostructure of a material.

**2. The mapping approach for the OSHS system.**

The distinctive temporal and spatial characteristics of the sound waves from the OSHS indicate in their similarity with the processes considered in the theory of dynamic chaos and nonlinear dynamic systems [1]. In that regards, we used some methods developed in this theory for the OSHS system. In order to study the spatial and temporal dynamics of an acoustical signal from the OSHS as well as the time series analysis we used a well-known and popular method for pseudo-attractor's reconstruction in the phase space. This method has been for the first time described by Packard [12] and has been proved in the Ruelle–Takens–Newhouse model [13]. The attraction of this approach is that it allows reconstructing the required pseudo-attractor by a time series of only one component $X(t)$ of the dynamic system trajectory. In accordance with the algorithm described in [13], the measured time dependency of a parameter responsible for system evolution is transformed to components of a vector in phase space of a single variable $X_{(m)}(t)$, ($m$ is embedding dimension):

$$X(t)=\{U(t),\ U(t+\tau),\ U(t+2\tau),\ldots,\ U(t+(m-1)\tau)\}. \qquad (1)$$

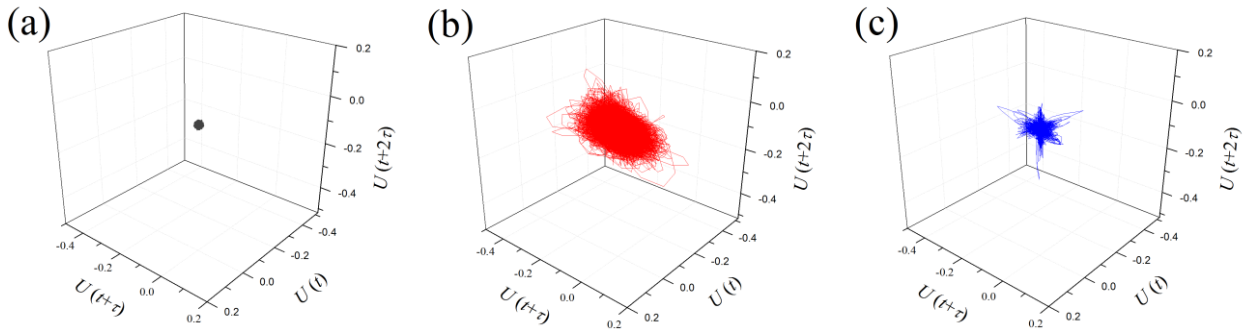

Fig.1. Phase portraits of a parameter $U(t)$ for consecutive OSHS stages ($m = 3$) in the same scale: (a) the charging material, (b) OSHS start-up, (c) the OSHS wave.

The time series (1) for the OSHS system was formed by discrete values of intensity $U$ of the sound wave generated by physical and chemical processes in the system. The parameter $U$ was measured with a digitization step of $\Delta t = 23\ \mu s$. The duration of the resulting audio files ranged in an interval from 3 to 11 seconds. Each of these vectors was calculated with a time delay $\tau$. This value is taken to be the time value whereby the autocorrelation function of the discrete sequence (1) goes to zero (or has the first minimum):

$$A(\tau) = \frac{1}{T}\sum_{t=0}^{T-1}(U_t - \langle U \rangle)(U_{t+\tau} - \langle U \rangle) \to 0. \qquad (2)$$

As we are aware, the role of the function $A(\tau)$ is not confined to that only. This function reveals interplay between random variables from the same time realization. Thus, it allows identifying transient processes and hidden periodicity.

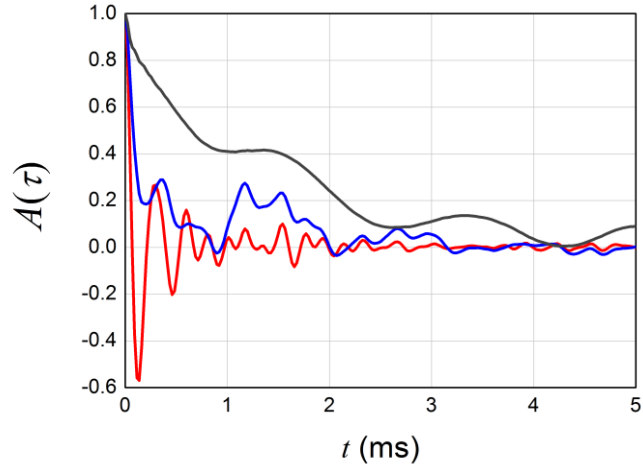

Fig.2. Autocorrelation functions of time series $U(t)$ for the modes presented in Fig.1 (in the same color scale).

A qualitative analysis of the different OSHS stages in the set of samples under study has shown the process of time evolution from a structureless set of trajectories (for the acoustic background and the charging material: the mixing of reagents and the quasi-isothermal reaction) to strong development forms of attractors (for the later stages of OSHS). An established fact in all the samples studied is that after the excitation of a system by an external thermal pulse, the phase space volume increases manifold. At this stage of the process, time series of $U(t)$ are characterized by non-Poisson statistics, i.e. the ratio of variance to the average value $\langle U^2 \rangle - \langle U \rangle^2 / \langle U \rangle \gg 1$. It is related to change of crystal structures as the result of the process temperatures increasing. OSHS stages directly related to the "combustion" process show complex phase portraits that include several types of motion. Their interpretation requires additional aspects of discussion and we will turn to these later. The formation of distinctive (quasi-periodic) phase portraits was described by us in [5]. We attributed that to formation and distribution of stationary wave of exothermic reaction in the stoichiometric mixture of reagents.

In addition, the respective autocorrelation functions gave absolute values of the periods. In the described system, functions $A(\tau)$ similar to that in [5] showed different periodic components for the different "combustion" modes [Fig. 2; same color scale in both figures (Fig. 1, Fig. 2)]. The periodic component $A(\tau)$ becomes already obvious at the stage of mixing of reagents (Fig. 2, a dark grey curve) and it is about 1 millisecond. This value appears to be related to the size of percolating cluster that is characteristic to the system. This is a distance at which the system almost repeats an initial disturbance [8]. In that regard, the phase space volume increases in 2-3 times in relation to the initial value (the acoustic background). The periodic component persists as well as after the excitation of a system by an external thermal pulse. At the same time its value decreases. In this system, "burning" stages are characterized by periods from several tens to hundreds of microseconds. The cooling and formation of the micro- and macrostructure of a resultant product is reflected in the change of steady-state conditions $A(\tau)$ by the dumping as well as in shrinking of phase space to initial sizes.

**3. Power spectral density and correlation dimension.**

Combinations of the phase space method, correlation analysis and spectral analysis have enabled us to get relevant research of possible dissipation channels of an energy which is dissipated in the form of an acoustic wave. In that context, we wish to identify in principle the causal links between observable features in the system studied. It should be noted that correlation analysis, by itself, gives no that but it only shows statistical correlations. According to the Wiener-Khinchin theorem, Fourier transform of the autocorrelation function $A(\tau)$ for time-discrete processes corresponds to the signal power spectral density defined as the sum:

$$F(\omega) = \sum_{t=-\infty}^{\infty} A(\tau) \cdot exp(-i\omega t), \qquad (3)$$

Where $\omega \in (-\pi, \pi)$. This characteristic is appropriate for use when considering nonstationary processes because of randomness and independence of spectral component phases in different time series samples. Fig.3 demonstrates the power spectra for OSHS stages studied in the set of samples studied. These stages specific to the time dependences of the acoustic signal have also been monitored by a real-time video process. It can be seen the power redistribution initially from the low-frequency range to the intermediate one. As can be seen in the Fig.3, a signal is the most powerful in the frequency range $\omega = 2 \div 5$ kHz that corresponds to the main "burning" stages.

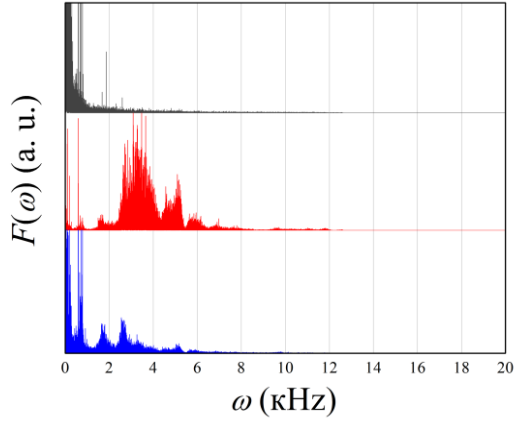

Fig.3. The spectral power density calculated from the functions $A(\tau)$ presented in Fig.2.

As has been indicated earlier, such a method in principle could provide the energy selectivity of devices based on OSHS processes [5]. The power spectral density approach shows the energy in the unit of time in the system and allows us in principle to distinguish quasi-periodic types of moves from chaotic ones. However, it has not any advantages in the analysis of chaotic regimes intrinsically because any information on the phase is lost. It is important information to the characterization of a strange attractor behavior [14]. It is known that the power spectral density has close connection with metric characterization of attractors, in particular the KS entropy that is its integral characteristic [15, 16]. The analysis of the quantitative data of attractors is traditionally starts with so called correlation dimension $D_c$ that is a quantitative characteristic of complexity (or quantitative estimation of self-similarity) of attractors. It can also serve as a measure of the internal structural order of an object as it is due to correlations between its structural elements. The procedure of this value estimation is described in more detail in [16]. The calculation of the $D_c$ value is reduced to that for the correlation sum:

$$C(r) = \frac{2}{n(n-1)} \sum_{i=0}^{n-2} \sum_{j=i+1}^{n-1} H(r - |U_i - U_j|), \qquad (4)$$

Where $H(x) = \begin{cases} 0, x \leq 0 \\ 1, x > 0 \end{cases}$ is the Heaviside step function for all pairs of $i$ and $j$, $n$ is the number of points $U_{i,j}$. $|U_i - U_j|$ is a distance between neighboring points in $m$-dimensional phase space, $r$ is the size of boxes covering the attractor (phase-space decomposition) exceeding the distance between two neighboring points. $C(r)$ increases with $r$ and has a saturation value finally. For the embedding dimension $m$ that is greater or equals greater or equal the topological dimension of the attractor the dependence (4) holds:

$$C(r) \propto r^{D_c}, \tag{5}$$

Where $D_c$ is the desired correlation dimension of the attractor. Having taken the logarithm of the expression (4), one receives a linear part of this dependence. The slope of that estimates $D_c$. The value $m_{max}$, where the slope does not change, corresponds to the true value of $m$. When the dependence $D_c(m)$ has clearly demonstrated a plateau in a wide range of distances, this plateau provides an estimate of the correlation dimension. The dynamics of steady-state processes must be small size, while transients can be multidimensional or even infinite-dimensional dynamical systems [1]. The correlation sum can be also used as a tool for recognizing dynamic irregularities arising from internal properties of an attractor [ib]. The dependence $D_c(m)$ calculated from (5) for the initial stage of OSHS (before the heat pulse initiation) behaves in a linear manner, in the first approximation (Fig.4, dark grey symbols). This is characteristic of an ordinary random process [17]. The dependence presented in Fig.4 by blue symbols (the OSHS wave) shows the maximum deviation from linearity but that is not reaching in itself.

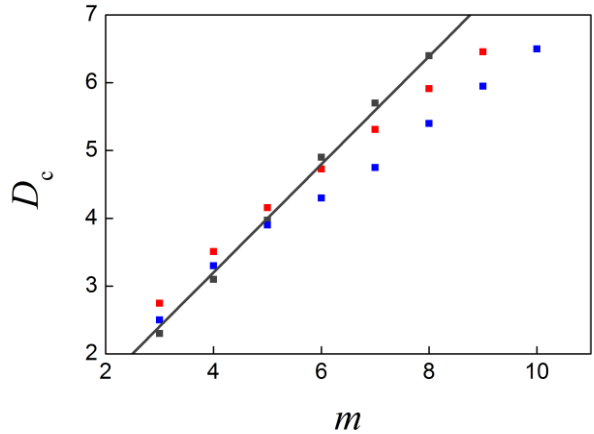

**Fig.4. The dependences $D_c(m)$ for 3 attractors presented in Fig.1 (in the same color scale).**

As a result, we cannot estimate the correlation dimension since the curves behave differently. This means appears to be that we deal with transition processes or it may be a consequence of a mixing in the system. That means statistical independence of different parts of a trajectory in phase space. In that case, it is hard to make an estimate of $D_c$ identically because various processes may be occurring at different values $m$. Thus, we face a question of how can we separate these various contributions?

## 4. Evaluation of the KS entropy and transition between regular types of the motion to chaotic one in the OSHS system.

The attractor reconstructed in phase space has fractal structure or self-similarity and it is quasi-periodic [1]. Self-similarity means that the process of feedback is going on repeatedly at different scales. It seems the simplest to find phase space scales at which a deterministic process is initiated. The KS entropy $h_{KS}$ mentioned above is a metric invariant of dynamic systems which defines whether there is a mixing in the system. It also allows determining the contribution of deterministic, regular and stochastic processes as well as average escape rate of trajectories on the attractor. It is usually estimated by Lyapunov characteristic exponents' $\lambda_m$ [16]. Their sum is equal to the entropy $h_{KS}$, by definition:

$$h_{KS} = \sum_m \lambda_m. \tag{6}$$

In our case, however, the calculation of exponents' $\lambda_m$ is hampered as we cannot uniquely determine the initial phase-space dimension as well as local instability. The phase-space trajectories seem to be belonging not to one but several processes. A positive $h_{KS}$ value characterizes the deterministic chaotic process. If this parameter increases without limit then we deal with a stochastic process. For $h_{KS} = 0$, the motion characteristics will be regular. In practice, for reliable estimating of $h_{KS}$ the value $h_2$ is used as a lower bound of $h_{KS}$. It is so-called correlation entropy or entropy per unit time (*correlation entropy*) that can be expressed by means of correlation sum [14]:

$$C(m,r) = r^{D_c} e^{-mh_2}, \tag{7}$$

It was shown in the work [18] that the concept of entropy per unit time can be used for a wide variety of processes including both deterministic and stochastic ones. The same problem is a subject of the review [19] that suggests a method of analysis of deterministic dynamic systems for large values of *m*. In this context, if the dependence $h_2(r)$ has a characteristic plateau then this behavior meets the criteria for deterministic chaos. It should be noted that in the present case, we are more interested not in convergence to specific positive value of $h_2$ but in essentially different values of this parameter for the various steps in the processes at a constant *m*. Since the parameter $h_2$ depends on a distance between trajectories *r*, one can attempt to define a contribution from the different processes due to its growth within a certain range of *r* at a given embedding dimension *m*. Shown in Fig.5 are the $h_2(r)$ dependences for 3 attractors (see Fig.1) at different $m$. We draw attention to the fact that the dependences for the charging material (dark grey curves) which are illustrating processes of solid-phase interaction of crystals and initial cracking do not have specific features for the presented values of $m$. Here, the entropy curves display conventional exponential decay. A noticeable non-monotonicity appears in the charts for "combustion" modes (red and blue curves). In the "mid-course" mode of "combustion" [$m = 6$, 7, Fig.5 (blue curve)], our system shows the picture similar to the one described in [18]. The characteristic plateau gives the value of $h_2 \approx 0.04$.

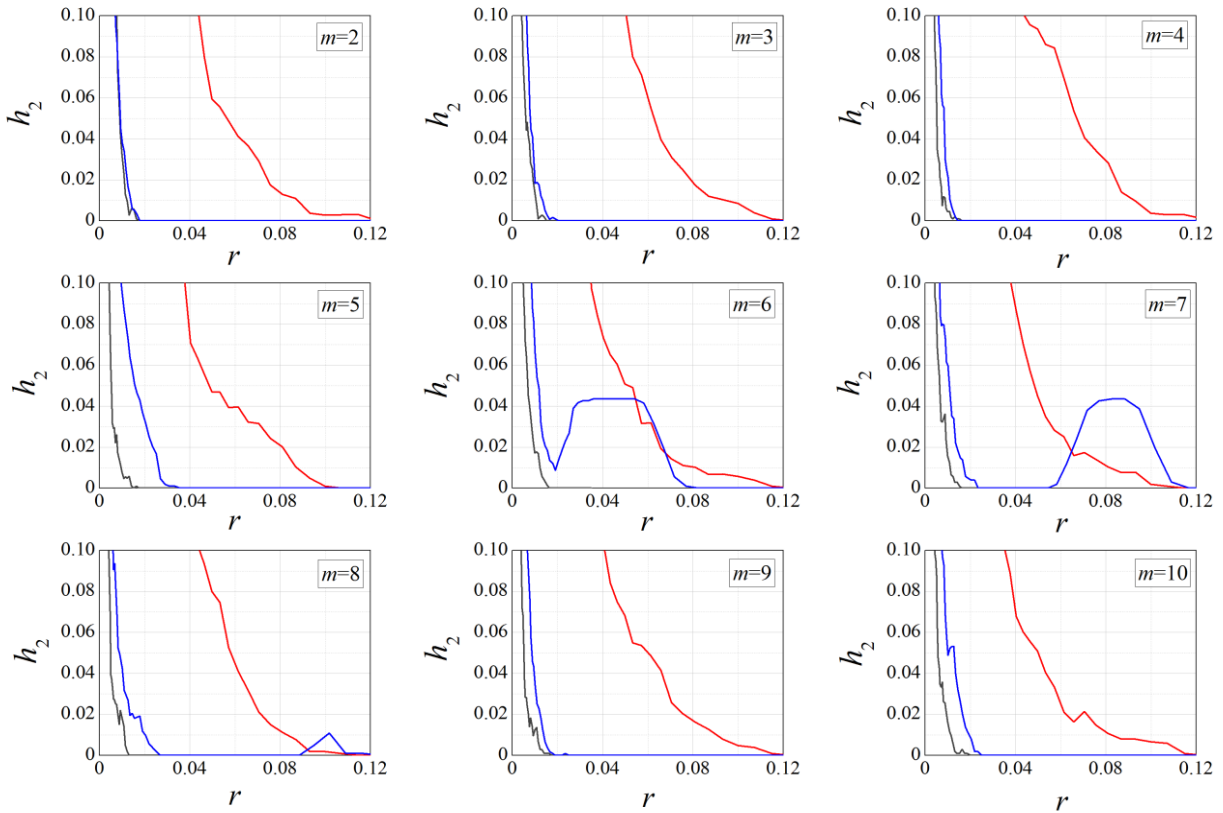

Fig.5. The dependences of $h_2(r)$ for 3 attractors ($m = 2 \div 10$) presented in Fig.1 (in the same color scale).

The dependences $h_2(r)$ represent the basis for evaluating the scale of $r$ when the deterministic process in the OSHS dynamics is beginning. As the work [18] argues, the higher the noise amplitude, the greater values of $r$ at which the plateau arises. That is the behavior can be observed in the mode "OSHS mid-course". In addition, overlapping effects in something called "mixed" phase space can be seen in this mode: red and blue curves in full ($m = 6$) or in part ($m = 7$) overlap at the scales $r = 0.03 \div 0.1$. So, the "combustion" stages have both chaotic and regular contributions that may occur simultaneously within the same scale of distances in the phase space of the acoustical response of system. We see the entropy per unit time serves here as an indicator of the transition from regular motion to chaotic one and back when going from one scale of *r* to another, and it can therefore be an indicator of the "combustion" efficiency.

**Conclusions**

In this paper, an analysis of spatial-temporal dynamics of the OSHS sound-wave intensity is used in order to evaluate the degree of its randomness and for identifying common factors in its individual stages. Autocorrelation functions of OSHS stages differ in appearance: from monotonous to oscillatory type. The lowest period of the oscillations seems to be related to the minimal size of percolating cluster that correlates with the mixture particle size. The initial step of "combustion" results in the multiple increasing in the phase space volume caused by formation and evolution of a non-stationary wave of exothermic reaction. Transition of the wave to the stationary mode or its change (when there is more than one stationary wave) resulting in stability of the phase space volume. The emission of sound waves associated with the processes of formation and shift of solid or liquid reaction environments that may occur as the result of deformation, friction and the crystal decrepitating at thermal expansion in solids or the bubble formation and their coalescence in liquids. Power spectra of OSHS stages allowed their characteristic frequency ranges. The analysis of metric characteristics of the OSHS attractors (correlation dimension and KS-entropy) revealed the transient nature of processes in the system and several types of movement simultaneously. KS-entropy evaluations found a deterministic phase as well as the transition to a regular phase in the solid-phase "combustion" process. The calculation of entropy in unit of time for various scales of *r* in the phase space of the control parameter *U* showed common characteristics when modifying time series structures (scale invariance). The value of that approach is that the study of entropy per unit time, depending on a runaway rate of trajectories in phase space at different values of embedding dimension, basically it is possible to answer on a question: at which value of *m* the efficiency of the mixture "combustion" process is highest? It could provide a basis for developing the theory having predictive force. At the same time, a plateau length in the dependence $h_2(r)$ shows compactness of the mapping in a simple and accessible format. The resulting qualitative picture of phase-space representation of the system studied requires enhanced analysis of such systems to reveal the possible statistically identical data. This would have the advantage of the study on the role of the percolation mechanism in OSHS transition processes.

**Acknowledgements**

The work has been supported by the Presidium of the Russian Academy of Sciences, program №0091-2015-00572017.

**References**

[1] H.G. Schuster, W. Just, Deterministic Chaos: an introduction, Physik Verlag, Weinheim, 248 p, 1984.

[2] M. Henon, "A two-dimensional mapping with a strange attractor", Comm. Math. Phys. **50** (1), 69-77 (1976).

[3] A.S. Pikovsky, "A dynamical model for periodic and chaotic oscillations in the Belousov-Zhabotinsky reaction", Physics Letters **85A**, 13 (1981); A. Cohen and I. Procaccia, Phys. Rev. **A 31**, 1872 (1985).

[4] K. Mainzer. "Complexity in chemistry and beyond: interplay theory and experiment, edited by C. Hill and D.G. Musaev, Springer, 2008.

[5] E.G. Klimchuk, A.L. Parakhonsky, "Acoustic investigations of organic SHS", *in Zel'dovich memorial*: *Accomplishments in the Combustion Science in the Last Decade*, Eds. A.A. Borisov, S.M. Frolov, Moscow: Torus Press, 2015, vol. 2, pp. 186-190.

[6] A.G. Merzhanov, "The chemistry of self-propagating high-temperature synthesis", Journal of Materials Chemistry **14**, 2004, pp. 1779-1786.

[7] E.G. Klimchuk, "Autowave exothermic organic synthesis in the mixes of organic solids", Macromolecular symposia, 2000, v.160, p.107.

[8] A. S. Mukasyan, S. Hwang, A. E. Sytchev, A. S. Rogachev, A. G. Merzhanov and A. Varma, "Combustion wave microstructure in heterogeneous gasless systems", Combustion Sci. Technol. **115**, 335 (1996).

[9] V. V. Brazhkin, A. G. Lyapin, S. V. Popova, and R. N. Voloshin, "New Types of Phase Transitions: Phenomenology, Concepts, and Terminology", in New Kinds of Phase Transitions: Transformations in Disordered Substances, edited by V. V. Brazhkin. S. V. Buldyrev, V. N. Ryzhov, and H. E. Stanley (Kluwer, Dordrecht, 2002) p. 15.

[10] E.G. Klimchuk "Acoustic effects in a wave of organic SHS, their diagnostics and interpretation". XII International symposium on self-propagating high temperature synthesis (SHS-2013): book of abstracts, South Padre Island, TX, USA, 37-38.

[11] E.G. Klimchuk at all, "Regularities of SHS in the solid-phase 8-oxyquinoline -chloramine B system". Russ. Chem. Bull. **48** (12), 1999, p.2245.

[12] N.H. Packard, J.P. Crutchfield, J.D. Farmer, and R.S. Shaw, Phys. Rev. Lett. 45 (9), 712 (1980).

[13] F. Takens, "Detecting strange attractors in turbulence". Lecture Notes in Mathematics, Berlin: Springer-Verlag, 898 p, 1981.

[14] J.-P. Eckmann and D. Ruelle, Rev. Mod. Phys. 57, 617 (1985).

[15] A.N. Kolmogorov, "New metric invariant of transitive dynamical systems and automorphisms on Lebesgue spaces", DAN SSSR. - 1958, v. 119, pp. 861-864.

[16] A. Yu. Loskutov, "Fascination of chaos", UFN, 2010, v. 180, 12, pp. 1305-1329.

[17] P. Grassberger, I. Procaccia, Phys. Rev. **A 28**, 2591 (1983).

[18] P. Gaspard and X-J. Wang, Physics Reports (Review Section of Physics Letters) 235 (6), 291-343, (1993).

[19] H. Kantz and T. Schreiber, Nonlinear time series analysis, Cambridge University Press, 2nd Edition, Cambridge, 371 p. (2004).